\documentclass[preprint,resetfootnote]{aastex7}
\usepackage[T1]{fontenc}
\usepackage[utf8]{inputenc}

\renewcommand{\t}{\theta}

\shorttitle{A Quadruply Lensed LLAGN in DELVE}
\shortauthors{Schechter et al.}

\graphicspath{./}

  \usepackage[T1]{fontenc}
  \usepackage[utf8]{inputenc}

\usepackage{hyperref}  

\usepackage[T1]{fontenc}
\usepackage[utf8]{inputenc}

\DeclareUnicodeCharacter{0327}{\textendash}
\DeclareUnicodeCharacter{0301}{\textendash}

\begin{document}
\input epsf


\title{The DELVE Quadruple Quasar Search I.  A Lensed Low Luminosity AGN}

\correspondingauthor{Paul L. Schechter}
\email{schech@mit.edu}

\author[0000-0002-5665-4172]{Paul L. Schechter}
\email{schech@mit.edu}
\affiliation{MIT Kavli Institute and Department of Physics, 77 Massachusetts Ave, Cambridge, MA, 02139,USA}

\author[orcid=0000-0001-6116-2095]{Dominique Sluse}
\email{dsluse@ulg.ac.be}
\affiliation{STAR Institute, Quartier Agora - All\'ee du six
Ao\^{u}t,  
19c B-4000,  Li\'ege, Belgium}

\author[orcid=0000-0002-6779-4277]{Erik A. Zaborowski}
\email{zaborowski.11@osu.edu} 
\affiliation{Department of Physics, The Ohio State University, Columbus, OH 43210, USA}
\affiliation{Center for Cosmology and Astro-Particle Physics, The Ohio State University, Columbus, OH 43210, USA}
\affiliation{Kavli Institute for Cosmological Physics, University of Chicago, Chicago, IL 60637, USA}

\author[orcid=0000-0001-8251-933X]{Alex Drlica-Wagner}
\email{kadrlica@uchicago.edu}
\affiliation{Fermi National Accelerator Laboratory, P.O. Box 500, Batavia, IL 60510, USA}
\affiliation{Department of Astronomy and Astrophysics, University of Chicago, Chicago, IL 60637, USA}
\affiliation{Kavli Institute for Cosmological Physics, University of Chicago, Chicago, IL 60637, USA}

\author[orcid=0000-0003-2456-9317]{Cameron Lemon}
\email{cameron.lemon@fysik.su.se} 
\affiliation{Institute of Physics, Laboratory of Astrophysics, Ecole Polytechnique F\'ed\'erale de Lausanne (EPFL), Observatoire de Sauverny, 1290 Versoix, Switzerland}

\author[orcid=0000-0000-0000-????]{Frederic Dux}
\email{duxfrederic@gmail.com}
\affiliation{Institute of Physics, Laboratory of Astrophysics, Ecole Polytechnique F\'ed\'erale de Lausanne (EPFL), Observatoire de Sauverny, 1290 Versoix, Switzerland}
\affiliation{European Southern Observatory, Alonso de C\'ordova 3107, Vitacura, Santiago, Chile}

\author[orcid=0000-0003-0758-6510]{Frederic Courbin}
\email{frederic.courbin@epfl.ch}
\affiliation{Institute of Physics, Laboratory of Astrophysics, Ecole Polytechnique F\'ed\'erale de Lausanne (EPFL), Observatoire de Sauverny, 1290 Versoix, Switzerland}
\affiliation{ICC-UB Institut de Ci\'encies del Cosmos, University of Barcelona, Mart\'i Franqu\'es, 1, E-08028 Barcelona, Spain}
\affiliation{ICREA, Pg. Llu\'is Companys 23, Barcelona, E-08010, Spain}

\author[orcid=0000-0002-6315-3085]{Angela Hempel}
\email{ang.hmpl@gmail.com}   
\affiliation{Instituto de Astrofisica,  Universidad Andr\'es Bello,
Fern\'andez Concha 700, 7591538 Santiago de Chile, Chile}
\affiliation{Max-Planck
  Institute for Astronomy, K\"onigstuhl 17, D-69117 Heidelberg, Germany}

\author[orcid=0000-0001-7051-497X]{Martin Millon}
\email{martin.millon@hotmail.fr}
\affiliation{Institute of Physics, Laboratory of Astrophysics, Ecole Polytechnique F\'ed\'erale de Lausanne (EPFL), Observatoire de Sauverny, 1290 Versoix, Switzerland}
\affiliation{Kavli Institute for Particle Astrophysics and Cosmology, Stanford University, Stanford, CA 94305, USA}
\affiliation{Department of Physics, Stanford University, 382 Via Pueblo Mall, Stanford, CA 94305, USA}

\author[orcid=0000-0002-8460-0390]{Tommaso Treu}
\email{tt@astro.ucla.edu} 
\affiliation{Department of Physics and Astronomy, University of California, Los Angeles, CA 90095, USA}

\author[orcid=0000-0002-5279-0230]{Raul Teixeira}
\email{raul.teixeira@duke.edu}
\affiliation{Department of Astronomy and Astrophysics, University of Chicago, Chicago, IL 60637, USA}

\author[orcid=0000-0002-6904-359X]{Monika Adam\'ow}
\email{madamow@illinois.edu}
\affiliation{Center for Astrophysical Surveys, National Center for Supercomputing Applications, 1205 West Clark St., Urbana, IL 61801, USA}

\author[orcid=0000-0003-4383-2969]{Clecio R. Bom}
\email{clecio@debom.com.br} 
\affiliation{Centro Brasileiro de Pesquisas F\'isicas, Rua Dr, Xavier Sigaud 150, Rio de Janeiro, RJ, Brazil}

\author[orcid=0000-0002-3690-105X]{Julio A. Carballo-Bello}
\email{jcarballo@academicos.uta.cl}
 \affiliation{Instituto de Alta Investigaci\'on, Universidad de Tarapac\'a, Casilla 7D, Arica, Chile}

\author[orcid=0000-0001-6957-1627]{Peter S. Ferguson}
\email{peter.ferguson@wisc.edu} 
\affiliation{Department of Physics, University of Wisconsin-Madison, Madison, WI 53706, USA}

\author[orcid=0000-0002-4588-6517]{Robert A. Gruendl}
\email{gruendl@illinois.edu} 
\affiliation{Center for Astrophysical Surveys, National Center for Supercomputing Applications, 1205 West Clark St., Urbana, IL 61801, USA}

\author[orcid=0000-0001-5160-4486]{David J. James}
\email{djames44@gmail.com}
\affiliation{ASTRAVEO, LLC, PO Box 1668, Gloucester, MA 01931, USA}
\affiliation{Applied Materials, Inc., 35 Dory Road, Gloucester, MA 01930, USA}

\author[orcid=0000-0002-9144-7726]{Mart\'{i}nez-V\'azquez}
\email{clara.martinez@noirlab.edu} 
\affiliation{NSF's NOIRLab, 670 N. A'ohoku Place, Hilo, HI 96720, USA}

\author[orcid=0000-0002-8093-7471]{Pol Massana}
\email{pol.massana@noirlab.edu}
\affiliation{NSF's NOIRLab, Casilla 603, La Serena, Chile}

\author[orcid=0000-0003-3519-4004]{Sidney Mau}
\email{smau@stanford.edu} 
\affiliation{Department of Physics, Stanford University, 382 Via Pueblo Mall, Stanford, CA 94305, USA}
\affiliation{Kavli Institute for Particle Astrophysics and Cosmology, Stanford University, Stanford, CA 94305, USA}

\author[orcid=0000-0001-9649-4815]{Burçin Mutlu-Pakdil}
\email{Burcin.Mutlu-Pakdil@dartmouth.edu}
\affiliation{Department of Physics and Astronomy, Dartmouth College, Hanover, NH 03755, USA}

\author[orcid=0000-0002-8282-469X]{No\"elia E. D. No\"el}
\email{n.noel@surrey.ac.uk} 
\affiliation{Department of Physics, University of Surrey, Stag Hill Campus, Guildford, GU2 7XH, UK}

\author[orcid=0000-0002-6021-8760]{Andrew B. Pace}
\email{apace@virginia.edu} 
\affiliation{McWilliams Center for Cosmology, Carnegie Mellon University, 5000 Forbes Ave, Pittsburgh, PA 15218, USA}

\author[orcid=0000-0002-1594-1466]{Joanna D. Sakowska}
\email{j.sakowska@surrey.ac.uk}
\affiliation{Department of Physics, University of Surrey, Stag Hill Campus, Guildford, GU2 7XH, UK}

\author[orcid=0000-0003-1479-3059]{Guy S. Stringfellow}
\email{Guy.Stringfellow@Colorado.EDU}
\affiliation{Center for Astrophysics and Space Astronomy, University of Colorado Boulder, 389 UCB, Boulder, CO 80309, USA}

\author[orcid=0000-0002-9599-310X]{Erik J. Tollerud}
\email{etollerud@stsci.edu}
\affiliation{Space Telescope Science Institute, 3700 San Martin Dr, Baltimore, MD 21218, USA}

\author[orcid=0000-0003-4341-6172,sname='Vivas']{A.~Katherina~Vivas}
\affiliation{Cerro Tololo Inter-American Observatory/NSF's NOIRLab, Casilla 603, La Serena, Chile}
\email{kathy.vivas@noirlab.edu}

\author[orcid=0000-0001-6455-9135]{Alfredo Zenteno}
\email{alfredo.zenteno@noirlab.edu}
\affiliation{Cerro Tololo Inter-American Observatory/NSF's NOIRLab, Casilla 603, La Serena, Chile}







\begin{abstract}
  
A quadruply lensed source, J125856.3$-$031944, has been discovered
using the DELVE survey and WISE $W1 - W2$ colors.  Followup
direct imaging carried out with the Magellan Baade 6.5 m telescope
is analyzed, as is spectroscopy from the 2.5 m Nordic Optical Telescope.
The lensed image configuration is kite-like, with the major axis of
the lensing galaxy along the symmetry axis of the kite, and with the
faintest image at its tail.
Redward of $6000  \mbox{\normalfont\AA}$ the tail image is
strongly blended with the lensing galaxy. 
The Sloan $g$ direct imaging carried out with Magellan
permits deblending. 
As the lensed image configuration is nearly circular,
simple models give high predicted magnifications for all four
images.  The source's narrow emission lines at redshift $z=2.225$ and low intrinsic
luminosity qualify it as a Type 2 AGN.  The Magellan image shows
a substantial residual that suggests a second lensing galaxy.

\end{abstract}


\keywords{Strong gravitational lensing (1643), Seyfert Galaxies (1447),
  Quasars (1319)}


\section{Introduction} \label{sec:Introduction}

Quadruply lensed quasar systems are routinely used to study the
lensing galaxies \citep{Vegetti}, the quasars themselves \citep{Vernardos},
and the geometry of the universe through which their light propagates
\citep{treumarshall}.  But they are rare,
both because quasars are themselves rare and because they have only an
${\mathcal O}(10^{-4})$ chance of being quadruply lensed \citep{Oguri}.
Worse yet, several observational considerations make them 
difficult to identify, the most important of which is their small
angular extent on the sky, $\sim 1\arcsec$.

Here we report the discovery of the first confirmed quadruply lensed
AGN found in the DELVE Quadruply Lensed Quasar (henceforth DELQQ)
survey, carried out in the DELVE footprint \citep{Alex},
using the Dark Energy Camera (DECam)
at the Victor M.\ Blanco 4 m telescope \citep{Flaugher}.  J125856.3$-$031944 (henceforth
DELQQ 1258$-$0319) presents several challenges.  Foremost among these,
the lensing galaxy overlaps with several of the images, making
deblending difficult.  Using only the original survey data, we could
not thoroughly convince ourselves that the system was indeed
quadruple, and not a ``naked cusp'' configuration like that of
J0457$-$7820 \citep{Lemon},  because the faintest quasar
image  is too close to the lensing galaxy.  

In higher resolution followup data there appears to be a second
lensing galaxy that lies close to two of the brighter images,
complicating modeling of the system.  The image configuration is
nearly circular, so small uncertainties in the image positions produce
large changes in the resulting models \citep{Falor}.

As deeper ground-based surveys are carried out,
{most recently DELVE \citep{Alex} and soon, the Legacy Survey of Space
  and Time \citep{LSST}},
quadruple quasars with
increasingly faint images relative to the galaxies that lens them will
become more prevalent and their modeling may be similarly
challenging.

\begin{figure}[b]
\hskip 1.75truein  
  \includegraphics[width=0.50\textwidth]{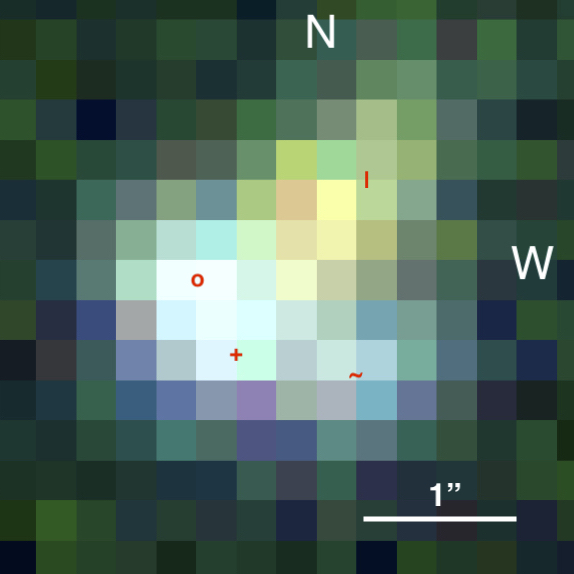}\label{fig:rgbORIG}  
  \caption{A color composite representation of DELQQ 1258$-$0319 taken with DECam on the
    Blanco Telescope, using  DES $r, i,$ and $z$ exposures (for
    blue, green and red) taken as part of the DELVE
    survey.  The approximate positions of four images subsequently labelled
  $A, B, C,$ and $D$ are shown in red with the symbols  $\circ$, +, $\sim$, and $|$, respectively.}
\end{figure}

\section{Discovery}\label{sec:Discovery}

\subsection{The DELQQ survey: selection}\label{subsec:DEL4Q}

Our survey, which we call DELQQ, covered the 6000 square degrees of
the first DELVE data release \citep{Alex}.  Details of the
survey are given in Appendix \ref{app:overview}. Candidate quasars with
color  $W1 - W2 > 0.7$ and $W1 < 15.5$ as measured by
the WISE \citep{Wright} satellite were chosen following the
color criterion proposed by \citet{Stern},
for a total of 150,000 sources.  
Square FITS cutouts, 45 pixels ($\sim 11\farcs85$) on a side,
were then
generated  for those candidates from the DELVE survey $g, r, i$ and $z$ images.

A still-evolving computer program, {\tt trifurcator}, first described
in \citet{Schechter_2018}, was used to split sources into three
components, which were classified as pointlike, $\mathcal P$, galaxy-like,
$\mathcal G$, or ambiguous $\mathcal Q$.  Systems for which none of the
components were pointlike were eliminated.

Cutouts in all four filters were visually inspected 
{by the first author} 
for systems in
which two components (one of which was pointlike) had near-identical
$g-r$, $r-i$ and $i-z$ colors, excluding those whose colors were
galaxy-like.  Roughly 1500 systems were visually inspected.
{\tt Trifurcator} is described in greater detail in Appendix \ref{app:trifurc}.

Five quadruply lensed quasars with $W1-W2 > 0.7$ had previously been
identified in our DELQQ sample
{of which {\tt trifurcator} recovered three:
  J1131$-$4419 \citep{Delchambre},
  J1134$-$2103 \citep{Lucey}, and
  J1131-1231 \citep{Sluse}.
}
The largest image separations for the
two systems not recovered were $0\farcs66$
for B1113$-$0641 \citep {Blackburne} and $10\arcsec$
 for J1651$-$0417 \citep{SternGral}
which were either too close to be resolved by DECam
or too widely separated for our search.  One known quad J1606$-$2333 {\citep{Gaia}}
lacked an $r$
exposure.  As of submission of this paper, we were unaware of any
other quadruply lensed systems discovered within the DELVE footprint.

\subsection{A candidate}\label{subsec:candidate}

Only one candidate emerged as a possible quadruply lensed quasar
system, DELQQ J1258$-$0319.  A color image generated from
the DELVE $r, i,$ and $z$ exposures is shown in Figure \ref{fig:rgbORIG}.

{\tt Trifurcator} split the $r$ exposure into three components: a
pointlike ${\mathcal P}$ component to the SW, marked by a red ``$\sim$''
in Figure \ref{fig:rgbORIG}, and two extended, galaxy-like ${\mathcal G}$ components,
one centrally located (a blend of an apparent lensing galaxy and an image marked
by the symbol ``$|$'') and one toward the East (a blend  of two images of the
source marked by the symbols ``$\circ$'' and ``$+$'').  The colors of this blend
agree with those of the pointlike object 
with an rms of 0.14 magnitudes, after allowing for a modest slope due
to micro-lensing, extinction and PSF mismatch.  They were flagged as
possible components of a quadruply lensed quasar.  The central galaxy-like
component has much redder colors.

The program assigned a DELVE $g$ magnitude of 22.68 to the pointlike image and
$g-r$, $r-i$ and $i-z$ colors of 0.373, 0.279, and 0.291 respectively.
The latter are useful for establishing achromaticity but less so for
deriving quasar properties, as the exposures spanned five observing
seasons.

While it was possible to split the central component into two point
sources, one of which would be a fourth quasar image, we could not
persuade ourselves that the data justified this.

A program called {\tt clumpfit}, described in Appendix \ref{app:clumpfit},
was used to fit five 
point sources simultaneously.  Though the decomposition of the two
${\mathcal G}$ images into point sources did give positions that were in
qualitative agreement with expectations for a quadruply lensed
source, reasonable alternative choices of possible PSFs produced large deviations
from these positions.

\subsection{Prior detections}\label{subsec:Prior}

Prior to its detection in DELVE, DELQQ J1258$-$0319 had been catalogued as
a source both in the Pan-STARRS catalogue \citep{PanSTARRS} and the VLASS
\citep{VLASS}.
Its non-detection in Gaia DR2 
indicates that it is substantially fainter than the large majority of
known quadruply lensed quasars.

\section{Followup MPIA-WFI imaging}\label{sec:WFI}


Four Cousins $R_c$ followup exposures of 320 s each were obtained
with the Wide Field Imager (WFI) of the MPIA 2.2 m telescope on La
Silla in seeing of roughly $0\farcs64$.
Notwithstanding the good seeing and pixels of $0\farcs238$,
there was no clear
separation of the lensing galaxy and the faintest of the quasar images.
Attempts by two of the authors to determine the {ellipticity} of the
lensing galaxy using differing techniques yielded substantially
different results.

\section{Followup NOT spectroscopy}\label{sec:NOT}

\begin{figure*}
\includegraphics[width=1.00\textwidth]{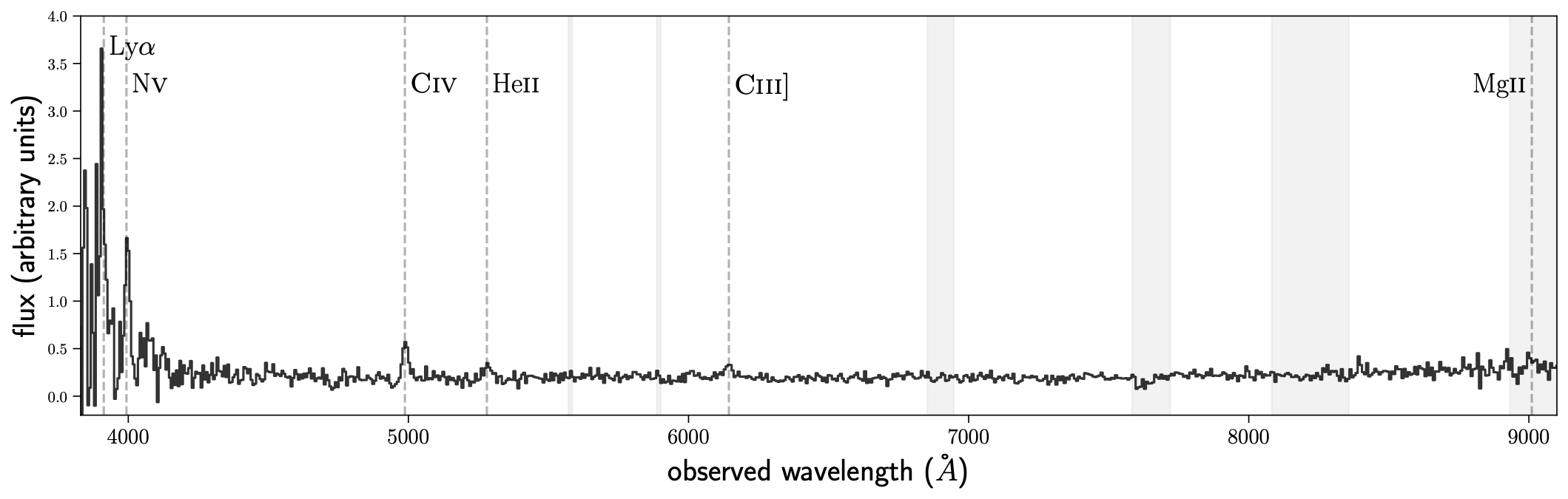}
\caption{
NOT spectrum of the brightest quasar component $A$,
extending from 3800 to 9000 $\mbox{\normalfont\AA}$.  The high
ionization lines of N$\;$V, C$\;$IV and He$\;$II seen at
$z = 2.225$ are narrow, with widths ranging from 1300-2000 km/s,
suggesting a Type 2 AGN rather
than a quasar.
}
\label{fig:spectrum}
\end{figure*}

\begin{figure*}[b!]
\hskip 0.75truein  
\includegraphics[width=0.80\textwidth]{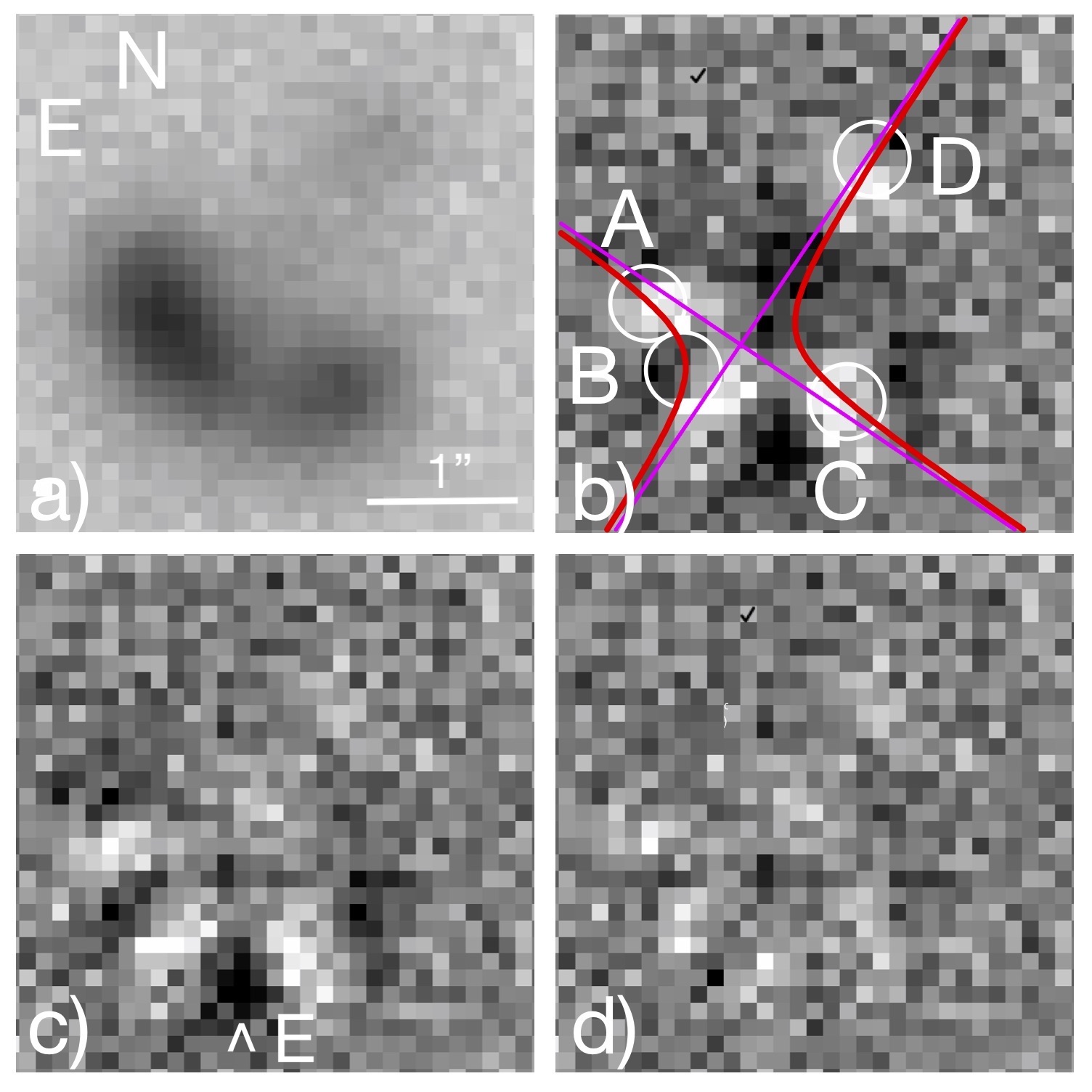}  
  \caption{(a) DELQQ 1258$-$0319 observed with a Sloan $g$ filter for
    900s with the IMACS f/4 camera on the Magellan Baade telescope.
    A  negative grayscale is used to highlight the low surface brightness of the
    central lens, which is substantially fainter than the four images surrounding it.
    (b) Four point sources, $A, B, C \& D$, have been fitted and
    subtracted.  A rectangular hyperbola (red) has been drawn through
    the fitted positions.  According to Witt (1996), if the potential
    is elliptical, the hyperbola's asymptotes (magenta) are parallel
    to its major and minor axes and the lensing galaxy must lie on the
    longer, primary branch of the hyperbola.  A faint source, that
    we take to be the lensing galaxy, is consistent with Witt's
    prediction.  (c)  A point source
    has been fitted to the lensing galaxy, which is too faint
    to permit determination of its extent.  White regions indicate
    oversubtraction of the point sources.
    (d) A substantial residual southeast of image $C$, which we call $E$,
    has been fitted by a point source and 
    subtracted.  The positions from this fit are shown in panel (b)
    as white circles.}
\label{fig:magdata}
\end{figure*}

A 40 minute spectrum was obtained with the Alhambra Faint Object
Spectrograph and Camera (ALFOSC) on the Nordic Optical Telescope (NOT)
on 2021 April 16.  The slit was oriented so
as to pass through the brightest of the four images and the lensing
galaxy.  A spectrum extracted at the position of the image ``$\circ$'' is shown in Figure 2.
Narrow emission lines are clearly visible, corresponding to a
redshift $z = 2.225.$


A sharp discontinuity in the galaxy spectrum is seen in the original
two-dimensional spectrum at 6000$\mbox{\normalfont\AA}$, suggestive of
the $H\&K$ break in early type galaxies.  In section
\ref{sec:Magellan} below we use astrometry obtained with Magellan's
Walter Baade Telescope {to force photometry on the original DELVE
data and} obtain colors for the lensing galaxy.  {The resulting}
photometric redshift {\bf is} consistent with this interpretation.


\section{Followup Magellan IMACS imaging}\label{sec:Magellan}

\subsection{2022 May}

Three 300 s exposures of DELQQ J1258$-$0319 in each of the Sloan $g$
and $i$ filters were obtained with the IMACS f/4 camera
\citep{Dressler} on the Baade 6.5 m telescope of the Magellan
Observatory on 2022 May 24, at a scale of $0\farcs111$ per pixel.
These sufficed to distinguish the fourth
image, ``$|$,'' from the much redder lensing galaxy, and to get positions
for each of them.  Attempts to determine the shape of the lensing
galaxy were inconclusive, as it blended with the three bright quasar
images.

\subsection{2024 July}\label{subsec:2024July}

Three IMACS 300 s exposures of DELQQ J1258$-$0319 in the Sloan $g$
filter were obtained
on 2024 July 6, a coaadition of which is shown in Figure \ref{fig:magdata}(a).
These had better seeing than the 2022 May images.  A nearby star 
($\alpha = 194.7527, \delta = -3.3168$)
was used to construct an empirical PSF, which was then fitted
to the four quasar images {\it and} to the lensing galaxy, again
using {\tt clumpfit}, described in Appendix \ref{app:clumpfit}.  As with
the earlier IMACS data, overlap between the galaxy and the
quasar images precluded determination of the shape or size of the lensing
galaxy.

Residuals obtained by subtracting five point sources from the stacked
image are shown in Figure \ref{fig:magdata}(b).  There is a bright
spot (which is dark on our inverted color map) between quasar images
$B$ and $C$, which we call $E$. Treating this as a sixth point source
gives the very much improved residuals shown at the same contrast in
panel \ref{fig:magdata}(d).  The positions of the four quasar images
showed appreciable shifts from panel \ref{fig:magdata}(c).

Fits of a singular isothermal elliptical potential (SIEP) model
to these new positions, described in section \ref{sec:Models} below,
produced a $\chi^2$  smaller by a factor of $(4.7)^2$  than did fits to the
positions of panel \ref{fig:magdata}(c).  We take this as strong evidence that
$E$ is a second lens rather than a fifth image of the source.  


The Magellan $g$ images were taken in non-photometric conditions.  The
DELVE catalogue mean magnitude for the empirical PSF star is $g =
20.322$ which gives $g=24.53 \pm 0.05$ for the lensing galaxy.  
The Magellan positions for the six point sources in Figure
\ref{fig:magdata}(c) were forced onto the original $r, i,$ and $z$ DELVE images,
using the DELVE PSFs, to give photometry for the lensing galaxy.
Combining these with the Magellan $g$ magnitude we assign
colors to the lensing galaxy of $g-r = 2.17$,
$r-i = 1.60$, and $i-z = 0.55,$ treating it as a point source.

We obtained a photometric redshift using {\tt Bayesian Photometric
Redshift} (hereafter BPZ) described by \citet{Benitez} and \citet{Coe}.
Prior distributions were derived from Hubble Deep Field North
galaxies.  We ran BPZ with the point-source fit apparent magnitudes
given in Table \ref{tab:main}.  We took the photometric
uncertainty to be $\pm 0.05$ mag in Magellan $g$
{which is the formal error reported from the {\tt clumpfit} covariance matrix
for objects $A$ - $E$ and $G$, with positions taken to be free}
We took it to be and $\pm 0.07, \pm 0.05$ and $\pm 0.04$ in $r, i,$ and $z$, respectively,
{which are the formal errors reported from the {\tt clumpfit} covariance
  matrices from the forced fits to the  individual DELVE cutouts.}

The best-fit redshift was found to be $z_{lens} =0.90 \pm 0.06$.
Multiple experiments were carried out exploring possible systematic
effects in the magnitudes and the effects of underestimating the
uncertainties,
{among them increasing our adopted errors by factors of two and four}.
In all but the most extreme cases the results were consistent with the best fit.

\begin{deluxetable}{crrrrrrrr}\label{tab:main}
\tablecolumns{9}
\tablewidth{720pt}
\tablecaption{DELQQ 1258$-$0319: Astrometric, Photometric and Modeled Quantities}
\tablehead{
  \colhead{~~image~~}    &
  \colhead{~~$\Delta$ R.A.~~} &  \colhead{~~$\Delta$ Dec.~~}  &
  \colhead{~~$g$~~} &
  \colhead{~~$g - i^*$~~} &
  \colhead{~~$F_g$~~} &
  \colhead{~~$\mu$~~} &
  \colhead{~~$\kappa^\dagger$~~} & \colhead{~~$\Delta t$~~} \\
  \colhead{ }    &
  \colhead{$\arcsec$} &  \colhead{$\arcsec$}  &
  \colhead{} &
  \colhead{} &
  \colhead{~~nJy~~} &
  \colhead{} &
  \colhead{} & \colhead{~~days~~} 
}
\startdata
 A &  0.96~~~ & -0.28~~~ & 22.13 & 0.91~~ & 1405 & 65.7 & 0.503 & ~2.52 \\
 B &  0.72~~~ & -0.93~~~ & 23.06 & 1.13~~ & ~964 &-190. & 0.526 & ~2.62 \\ 
 C & -0.37~~~ & -0.94~~~ & 22.46 & 1.21~~ & ~847 & ~8.7 & 0.467 & ~0.00 \\ 
 D & -0.54~~~ &  0.68~~~ & 23.02 & 1.64~~ & ~620 & -5.7 & 0.587 & ~24.0 \\
 E &  0.17~~~ &  1.10~~~ & 23.25 & 1.38~~ & ~501 &         &            &    \\       
 G & 0.00~~~  &  0.00~~~ & 23.52 & 2.74~~   &       &      &        &    \\
\enddata
\tablenotetext{$*$}{Colors assume a pointlike lens.  Extended flux from the
much redder lens contaminates the colors of $A$-$E$.
$^\dagger$For an SIEP, shear $\gamma =$ convergence $\kappa$.}
\end{deluxetable}

\section{Lens Models}\label{sec:Models}
We used Keeton's {\tt lensmodel} program \citep{Keeton} to model the
positions of the quasar images shown by the white circles in Figure
\ref{fig:magdata}(c) and given in Table \ref{tab:main}.  The lensing
galaxy was represented as a singular isothermal elliptical potential
(SIEP),
\begin{equation}
    \psi(b, q_{pot}) = b\sqrt{q_{pot}x^2 + \frac{y^2}{q_{pot}}} \quad .
\end{equation}
Here $\psi$ is the dimensionless projected potential and
$q_{pot}$ is the axis ratio of this potential (rather than that
of its surface mass density).\footnote{
For very round potentials, the underlying mass is roughly three times flatter
\citep{LuhtaruI}
} The lens strength  $b$ is measured in
arcseconds, and gives the Einstein ring radius in the limit of small
elllipticities.  These were treated as free parameters, as was the
position angle (P.A.) of the long axis.  The four image positions were assigned
uncertainties of $0\farcs01$.   Results from this fit are presented in the first
line of Table \ref{tab:keeton}.  The predicted magnification $\mu$ of image $C$ is also given.

\begin{deluxetable}     {lrrrrrr}\label{tab:keeton}
\tablecolumns{6}
\tablewidth{720pt}
\tablecaption{Model fits to image positions}
  \tablehead{
    \colhead{Model} &
    \colhead{$b_{siep}$} &
    \colhead{$b_{sis}$} &    
    \colhead{$\epsilon_{siep}$} &
    \colhead{P.A.$^*$} &
    \colhead{$\mu_C$} &
    \colhead{$\chi^2$}    \\
}
\startdata
SIEP            & $0\farcs97$  &                      & 0.020 & $-33^\circ$ & 34~~ & 87 \\
SIEP + SIS & $0\farcs93$ & $0\farcs03$ &0.045 & $-45^\circ$ &  9~~ &  4 \\
\enddata
\tablenotetext{$*$}{Position angles are measured from East to North}
\end{deluxetable}

The second line in Table \ref{tab:keeton} models the image positions with both an SIEP
and with a Singular Isothermal Sphere (SIS) at the position of point source $E$.
One additional parameter, $b_{sis}$,  has improved the mean
squared residual of the image positions by a factor of $(4.67)^2$.    Yet more remarkable,
the strength of the added SIS is a very small fraction, 1/30, that of the SIEP.

\citet{Falor} show that for asymptotically circular quads, the
magnification of the closest pair of images increases inversely as
their separation.  The SIEP centered on the lensing galaxy has an
ellipticity $\epsilon_{siep}$ of just 0.045, so even a small
perturbation can have a large effect on the close
pair.\footnote{\citet{LuhtaruI} find a typical ellipticity for the
potentials of quadruply
lensed quasars of 0.21} Our hypothesized perturber at position $E$ in
Figure \ref{fig:magdata}(c) pushes image $B$ closer to image $A$.
Falor and Schechter also found that for asymptotically circular quads, the
absolute magnifications for all four images vary inversely
with the ellipticity of the potential.

We have chosen to give the predicted magnification for image $C$ in
Table \ref{tab:keeton} because (as the leading image {in Table 
  \ref{tab:main}}) it is least subject to changes in the lensing
potential.  The predicted flux ratios {in Table \ref{tab:main}} $A/C $
and $B/C$ are 22 and 7, substantially higher than the observed ratios
of 1.7 and 1.1, respectively.

\citet{Weisenbach} have shown such large deviations from predicted
fluxes are unlikely to be the result of gravitational micro-lensing.
Nor do we think it is due to obscuration of image $C$ by dust,
as its $g-i$ colors are no different from those of $A$ and $B$.

We believe, instead, that the unusual flux ratios reflect a
shortcoming of our observations.  The lensing galaxy appears to be
elongated roughly along an axis from image $D$ to image $A$.  In
models that take the stellar axis ratio of the lens to be a free
parameter, it is strongly degenerate with the positions of those two
images.  Plausible changes in the image positions could produce
different modelled flux ratios, but would not be sufficiently large to
change the near-circular character of the potential.

\section{Quasar, LLAGN or Seyfert?}\label{sec:Seyfert}

Based on the narrow emission lines seen in Figure \ref{fig:spectrum},
\citet{Kachikian71} and \citet{Kachikian74} -- the first of these in
Russian and the second in English -- would have classified the source in
DELQQ 1258$-$0319 as a Seyfert II galaxy rather than a Seyfert I or a
quasar.  Subsequent work indicated that physics of the ``active
galactic nuclei" (AGNs) of such systems is somewhat decoupled from the
properties of the galaxies that host them.  In a review of "unified
models" \citet{Antonucci} presents a cartoon model in which Type 2 AGN
have narrow emission lines with FWHM ${\mathcal O}(1000)$ km/s, and
Type 1 AGN have broad emission lines with FWHM ${\mathcal O}(10000)$
km/s.  Such systems are classified as Seyferts only if the host
galaxy is observed, which may eventually happen if the system is
observed at sufficiently high spatial resolution and surface
brightness sensitivity.

The source in the DELQQ 1258$-$0319 system is not very red, as would
have been the case if dust were obscuring both its broad line region
and continuum emission from its accretion disk.  The high
magnifications derived from our models and the photometry from the
original DELVE images point to a source absolute magnitude $M_{AB}$
fainter than those of the low luminosity quasars in the SHELLQs survey
\citep{SHELLQs}.  It would therefore seem that the first quadruply
lensed AGN in the DELVE Quadruple Quasar survey does not qualify as
quasar.  As we do not, as yet, see evidence of the host galaxy,  the
source is classified as a low luminosity AGN -- ``LLAGN'' -- and not a
Seyfert.

\section{Conclusions}\label{sec:Conclusions}

With the benefit of hindsight, none of the challenges described in the
preceding sections is entirely surprising.  Many of the brighter
quadruply imaged quasars with wider separations have already been
discovered, particularly since the publication of Gaia DR2
\citep{Delchambre,Lemon}. The $W2 - W1 > 0.7$ criterion will not be
satisfied by systems for which the lensing galaxy contributes a
substantial fraction of the infrared light.  Lensed image
configurations often include faint images like $D$
(the highest saddlepoint of the light travel time)
lying relatively close to
the lensing galaxy, making
deblending more difficult.  Seyfert galaxies are more common than
quasars \citep{Willott}, and may account for an increasingly large
fraction of quadruply lensed sources in future surveys that push to
fainter apparent magnitudes.  And at fainter apparent magnitudes, it
may not be possible to deblend the targets into at least three images
in all four filters, as described in Appendix \ref{app:trifurc}.
Direct forward lens modelling of the pixels, rather than identifying
${\mathcal P}$, ${\mathcal G}$, and ${\mathcal Q}$ components of the
system in multiple filters, would permit quads to be identified in a
single filter.

The preceding sections reveal an as yet unquenched thirst for higher and
yet higher resolution, first to confirm that the AGN DELQQ 1258$-$0319
is in fact quadruply lensed and then to separate its multiple images
from the lensing galaxy, which would appear to have a companion.  One
might do yet better from the ground with ``lucky imaging''
\citep{Mackay}, but bright nearby stars usually needed for adaptive
optics are scarce.

Beyond that, one must get above the Earth's atmosphere.
In the time since this paper was first  submitted, DELQQ 1258$-$0319
was included in a successful HST ``bridge'' program which may
help resolve some of the outstanding questions.  

\begin{acknowledgments}
We warmly thank R. Gredel, H.-W. Rix and T. Henning for allowing us to
observe with the MPIA 2.2 m telescope.  We thank N. Zakamska for an
extended disquisition on the physics underlying the arcana of AGN
classification.  {We thank an anonymous referee whose suggestions
for clarification  precipitated the inclusion of the suplementary material 
in Appendices $A$ - $C$.}  This program has been supported in part by the Swiss
National Science Foundation (SNSF) and by the European Research
Council (ERC) under the European Union's Horizon 2020 research and
innovation programme (COSMICLENS: grant agreement No 787886).
M.\ M.\ acknowledges the support of the Swiss National Science
Foundation (SNSF) under grant P500PT203114.  T.\ T.\ acknowledges
support by the NSF through grants 1906976 and 1836016.

\end  {acknowledgments}

%


\vspace{5mm} \facilities{
Dark Energy Camera (DECam),
constructed by the Dark Energy Survey (DES) collaboration.
Funding for the DES Projects has been provided by 
the U.S. Department of Energy, 
the U.S. National Science Foundation, 
the Ministry of Science and Education of Spain, 
the Science and Technology Facilities Council of the United Kingdom, 
the Higher Education Funding Council for England, 
the National Center for Supercomputing Applications at the University of Illinois at Urbana-Champaign, 
the Kavli Institute of Cosmological Physics at the University of Chicago, 
the Center for Cosmology and Astro-Particle Physics at the Ohio State University, 
the Mitchell Institute for Fundamental Physics and Astronomy at Texas A\&M University, 
Financiadora de Estudos e Projetos, Funda{\c c}{\~a}o Carlos Chagas Filho de Amparo {\'a} Pesquisa do Estado do Rio de Janeiro, 
Conselho Nacional de Desenvolvimento Cient{\'i}fico e Tecnol{\'o}gico and the Minist{\'e}rio da Ci{\^e}ncia, Tecnologia e Inovac{\~a}o, 
the Deutsche Forschungsgemeinschaft, 
and the Collaborating Institutions in the Dark Energy Survey. 
Max Planck~2.2m(WFI), NOT(ALFOSC),
Magellan:Baade(IMACS).  Based on observations at Cerro Tololo
Inter-American Observatory, NSF’s National Optical-Infrared Astronomy
Research Laboratory (2019A-0305; PI: Drlica-Wagner), which is
operated by the Association of Universities for Research in
Astronomy (AURA) under a cooperative agreement with the National
Science Foundation.}





\appendix

\section{Description of the DELVE survey}\label{app:overview}

We quote verbatim the abstract of \citet{Alex} describing DELVE.

The DECam Local Volume Exploration survey (DELVE) is a 126-night
survey program on the 4\,m Blanco Telescope at the Cerro Tololo
Inter-American Observatory in Chile.  DELVE seeks to understand the
characteristics of faint satellite galaxies and other resolved stellar
substructures over a range of environments in the Local Volume.  DELVE
will combine new DECam observations with archival DECam data to cover
roughly $ \sim 15,000 \deg^2$ of high Galactic latitude ($|b| > 10\deg$)
southern sky to a $5 \sigma $ depth of $g,r,i,z \sim 23.5 $ mag.
In addition, DELVE will cover a region of 2200 $ \deg^2$
around the Magellanic Clouds to a depth of $g,r,i \sim 24.5 $ mag
and an area  $135  \deg^2$ around four Magellanic
analogs to a depth of $g,i \sim 25.5 $ mag.  Here, we present an
overview of the DELVE program and progress to date.  We also summarize
the first DELVE public data release (DELVE DR1), which provides
point-source and automatic aperture photometry for roughly $520$
million astronomical sources covering roughly $5000 \deg^2$ of the
southern sky to a $5\sigma$ point-source depth of $g{=} 24.3$ mag ,
$r{=} 23.9 $ mag,  $i{=}23.3$ mag, and $z{=} 22.8$ mag.
DELVE DR1 is publicly available via the NOIRLab Astro Data Lab science
platform.

Of particular relevance for the present work, the DELVE survey uses
{\tt PSFeX} \citep{Bertin} to produce a tabulated PSF at any desired
position. As illustrated in Figure 2 of \citet{Alex}, the median PSF
FWHM ranged from $\sim 1\farcs0$ in $z$ to $\sim 1\farcs25$ in $g$ and
was properly sampled by DECam's $0\farcs263$ pixels.

\section{How {\tt trifurcator} works}\label{app:trifurc}

A still-evolving program, {\tt trifurcator},
was used to select candidates from our WISE targets.  The program
deblends images in all four DELVE filters for each
of the WISE targets into 3 components using an algorithm drawn from
{\tt DoPHOT} \citep{Schechter_1993}, classifying the components as either
pointlike or extended, using the DELVE PSF as a reference.  It then
requires that at least one pointlike component has quasar-like colors
drawn from Table 3 of \citet{Richards} and that a second component
has colors identical to the pointlike component.  The second component
need not be pointlike, as it may be an unresolved blend of two quasar
images.  Systems for which none of the components are pointlike 
in any of the filters are eliminated.

If these criteria are satisfied in any one of the DELVE filters, the
three components are used to force photometry on the other DELVE
filters using the empirical PSF option in {\tt DoPHOT}. 

Cutouts in all four filters are then visually inspected for systems in
which two components (one of which was pointlike) had near-identical
$g-r$, $r-i$ and $i-z$ colors, {with a scatter less than 0.2 mag.
  If these colors were galaxy-like, the system was dropped from
  further consideration.}  Roughly 1500 systems were visually
inspected {by the first author}.

{In an experiment (to be reported elsewhere) conducted by two of the present authors,
two dozen  individuals with considerable lensing experience
were shown the image configurations for 3 of the 4 images of
known quasars and asked to mark the position of the fourth.  They did so
with high reliability, and in some cases identified reasonable alternative
positions.  The results instill a measure of confidence in the present
visual inspections.}

\section{How {\tt clumpfit} works}\label{app:clumpfit}
Like {\tt trifurcator}, {\tt clumpfit}  uses pieces of the program {\tt
  DoPHOT} \citep{Schechter_1993}, to obtain parameters for pointlike
and extended sources in a cutout.  Initial positions, and for extended
sources, initial shape parameters are specified and then
simultaneously adjusted to obtain a best fit.  One can  either let
the program solve for an analytic PSF or specify a PSF.  For the
original DELVE cutouts the DELVE PSF was used.  For the Magellan IMACS
data, a nearby star was used as a pixellated template.  The program has much in
common with {\tt GALFIT} \citep{Peng}.

\par

\bibliography{Schechter}{}
\bibliographystyle{aasjournal}



\end{document}